\documentclass[aps,prl,preprintnumbers,showpacs,twocolumn,nofootinbib]{revtex4}
\usepackage{graphicx}
\usepackage{equation}

\newcommand{\PRE}[1]{}       

\def\gappeq{\mathrel{\rlap {\raise.5ex\hbox{$>$}}
            {\lower.5ex\hbox{$\sim$}}}}
\def\lappeq{\mathrel{\rlap{\raise.5ex\hbox{$<$}}
            {\lower.5ex\hbox{$\sim$}}}}

\begin{document}
\renewcommand{\thefootnote}{\fnsymbol{footnote}}

\title{
\PRE{\vspace*{2.0in}}
Probing MeV Dark Matter at Low--Energy $e^+e^-$ Colliders
\PRE{\vspace*{.4in}}
}

\author{Natalia Borodatchenkova$^1$}
\author{Debajyoti Choudhury$^2$} 
\author{Manuel  Drees$^1$\PRE{\vspace{.1in}}
}
\affiliation{{}$^1$Physikalisches Institut der Universit\"at Bonn, Nussallee
  12, 53115 Bonn, Germany\\
{}$^2$Department of Physics and Astronomy, University of Delhi,
  Delhi 110007, India}
\begin{abstract}
  It has been suggested that the pair annihilation of Dark Matter particles
  $\chi$ with mass between 0.5 and 20 MeV into $e^+e^-$ pairs could be
  responsible for the excess flux (detected by the INTEGRAL satellite) of 511
  keV photons coming from the central region of our galaxy. The simplest way
  to achieve the required cross section while respecting existing constraints
  is to introduce a new vector boson $U$ with mass $M_U$ below a few hundred
  MeV. We point out that over most of the allowed parameter space, the process
  $e^+e^- \rightarrow U \gamma$, followed by the decay of $U$ into either an
  $e^+e^-$ pair or an invisible ($\nu \bar \nu$ or $\chi \bar \chi$) channel,
  should lead to signals detectable by current $B-$factory experiments.  A
  smaller, but still substantial, region of parameter space can also be probed
  at the $\Phi$ factory DA$\Phi$NE.

\end{abstract}

\maketitle

Within the context of Einsteinian gravity, evidence for the existence of Dark
Matter (DM) is overwhelming~\cite{silkrev}. Analyses of the cosmic microwave
background anisotropy, and other data on the large scale structure of the
Universe, have determined many cosmological parameters with unprecedented
precision~\cite{wmap}. Along with analyses of Big Bang nucleosynthesis (BBN)
\cite{bbn}, this data also show the DM must be largely non--baryonic. Since
neutrinos can contribute only a small fraction~\cite{silkrev}, this
strongly points towards the existence of an exotic, neutral, stable particle
$\chi$, with relic density $\Omega_{\chi}$ satisfying~\cite{wmap}
\begin{equation} \label{density}
\Omega_{\chi} h^2 = 0.113 \pm 0.0085 \quad ({\rm at} \ 1 \sigma)\, ,
\end{equation}
where  the scaled Hubble constant $h \simeq 0.7$~\cite{wmap}.

Dark Matter particles should clump due to gravitational attraction. At the
galactic center, their density might be so high that their annihilation into
lighter, known particles could lead to visible signals.  Final states
containing hard photons play a special role in this, since photons travel in
straight lines and are easy to detect.

Looking for an excess of hard photons from the center of our galaxy, the
INTEGRAL satellite indeed observed a large flux of photons with energy of 511
keV \cite{integral}. This sharp line can only come from the annihilation of
non--relativistic $e^+e^-$ pairs. However, most estimates of positron
production by astrophysical sources fall well short of the required flux
giving rise to the speculation \cite{boehm1} that the annihilation of light DM
particles $\chi$ into $e^+ e^-$ final states could be responsible for this
signal. While the mass $m_\chi$ can exceed $m_e$ substantially (since the
positrons produced in $\chi$ annihilation quickly lose energy through
scattering on neutral atoms), it has been pointed out~\cite{beacom} that
annihilation into $e^+e^-\gamma$ final states, in spite of being a higher-order
process, would over--produce MeV photons if $m_\chi > 20$ MeV, leading to
\begin{equation} \label{mass}
m_e \leq m_\chi \leq 20 \ {\rm MeV} \, .
\end{equation}
Recently, it has been argued \cite{ahn} that there exists evidence
for a non--astrophysical source of MeV photons, in which case the upper end of
the range (\ref{mass}) would be favored.

In order to produce the required flux of positrons, one needs an annihilation
cross section $\sigma_{\rm ann}(\chi \bar \chi \rightarrow e^+ e^-)$ 
\begin{equation} \label{eeann}
10^{-3} \ {\rm fb} \leq v \, \sigma_{\rm ann} \cdot
\left(  {m_\chi} / 1 \ {\rm MeV}\right)^{-2} \cdot \kappa \leq 1 \ {\rm fb}
\, . 
\end{equation}
Here $v$ is the relative velocity of the two $\chi$ particles in their cms
frame, and $\sigma$ is to be computed at $v \sim 10^{-3}c$. Note that models
of the galaxy fix the DM mass density; the number density, which enters
quadratically in the calculation of the positron flux, therefore scales as
$m_\chi^{-1}$. Finally, $\kappa = 1$ if $\chi$ is self--conjugate (i.e. a
Majorana particle), whereas $\kappa = 2$ if $\chi \neq \bar \chi$, since then
only half of all encounters of DM particles can lead to annihilation events.
To be on the safe side, in (\ref{eeann}) we have expanded the range given in
the original publication \cite{boehm1} by an order of magnitude in either
direction. This may be overly conservative \cite{asca}, since one here only
needs the DM density averaged over a significant volume, which is thought to
be better known than that right at the galactic center.

A cross section in the range (\ref{eeann}) implies that $e^+e^-
\leftrightarrow \chi \bar \chi$ reactions were in equilibrium down to
temperatures well below $m_\chi$. Since successful BBN requires a starting
temperature $T \geq 0.7$ MeV \cite{bbntemp}, it is safe to assume that $\chi$
particles indeed were in thermal equilibrium. Their relic density then turns
out to be \cite{kt} inversely proportional to the thermal average of their
total annihilation cross section into all final states containing only SM
particles:
\begin{equation} \label{omega}
 \Omega_{\rm DM} h^2 \propto \langle v \sigma_{\rm tot} \rangle^{-1}
 \,. 
\end{equation}
Given the mass constraint (\ref{mass}), to the leading order, annihilation is
possible only into $e^+e^-$ and $\nu \bar \nu$ final states; the first channel
must exist, since $\chi$ particles are supposed to annihilate into $e^+e^-$
pairs even today. Assuming $\chi$ forms (nearly) {\em all} DM, the relic
density resulting from eq.(\ref{omega}) must fall in the range
(\ref{density}). This constraint, interpreted using eq.(\ref{omega}), and
(\ref{eeann}) are compatible only if the present $\sigma_{\rm ann}$ is
strongly suppressed compared to that at decoupling. This is most easily
achieved \cite{bofa,boehm1,fayet} if $\chi \bar \chi$ annihilation proceeds
only from a $P-$wave initial state, in which case $v \, \sigma_{\rm ann}
\propto v^2$; note that $v^2 \sim 0.1$ when $\chi$ particles decoupled, while
$v^2 \sim 10^{-6}$ today.

In a renormalizable theory, some particle must mediate $\chi \bar \chi
\rightarrow e^+ e^-$ annihilation. The simplest possibility
\cite{boehm1,fayet} is to introduce a light spin--1 boson $U$ coupling to both
$e^+e^-$ and $\chi \bar \chi$ states. If $\chi$ is a Majorana spin--1/2
fermion or complex scalar, $\sigma_{\rm ann} (\chi \bar \chi \rightarrow f
\bar f)$ is given by \cite{fayet}
\begin{eqalignno} 
\label{cctoff}
v \sigma_{\rm ann} = \frac {\beta_f g_\chi^2 } {12 \pi s} 
& \left\{ \frac {(s-4m_\chi^2) \, 
                  \left[ s \Sigma_f + m_f^2 \left( 6 \Pi_f -\Sigma_f
                                 \right) \right]} 
                { (s-M_U^2)^2 + \Gamma_U^2 M_U^2 }
\right. \nonumber \\[-2ex] & \left.
+ \xi \; \left( \frac {m_f m_\chi} {M_U^2} \right)^2 \left( 3 \Sigma_f - 6
  \Pi_f \right) \right\} \, ,
\end{eqalignno}
where $\xi = 1 \, (0)$ for spinor (scalar), $\beta_f = \sqrt{1 - 4 m_f^2/s}$,
$\Pi_f = g_{f_L} g_{f_R}$ and $\Sigma_f = g_{f_L}^2 + g_{f_R}^2$, with
$g_{f_L}$ and $g_{f_R}$ being the left-- and right--handed $U f \bar f$
couplings. The $U \chi \bar \chi$ coupling $g_\chi$ is purely axial vector for
a Majorana $\chi$.  The first line in eq.(\ref{cctoff}) is a pure $P-$wave
contribution.  We discount, henceforth, a Dirac $\chi$ as it would, in
general, have a large $S-$wave contribution to $\sigma$ thereby making it
difficult to reconcile the constraints (\ref{density}) and (\ref{eeann}).

Since $\Gamma_U \ll M_U$ for realistic couplings, the usual \cite{kt}
non--relativistic expansion of the cross section in powers of $v$ breaks down
if $2 m_\chi \simeq M_U$ and the thermal averaging has to be done numerically
\cite{griest}. The strong velocity dependence of the cross section ($s - 4
m_\chi^2 \simeq v^2 m_\chi^2$ if $v^2 \ll 1$) in eq.(\ref{eeann}) also has to
be treated properly. For simplicity, we assume a thermal velocity
distribution, with a rather high temperature $T = 10^{-6} m_\chi$; smaller
temperatures, which are probably more realistic, would require larger
couplings, which would be easier to test at $e^+ e^-$ colliders.

\begin{figure}[h!]
\vspace*{-5mm}
\rotatebox{270}{\includegraphics[height=0.4\textwidth]{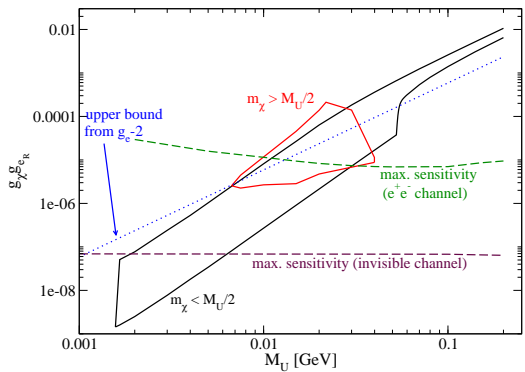}}
\vspace*{-3mm}
\caption{Parameter space for a Majorana $\chi$ with $g_{e_L} = g_\nu= 0$ and
  $g_\chi = 10 g_{e_R}$. In between the solid curves, $\chi$ has the correct
  relic density, and the correct cross section to explain the flux of 511 keV
  photons emerging from the galactic center; the red (black) curves correspond
  to $2 m_\chi < (>) M_U$. The dotted (blue) line indicates the upper bound on
  $g_{e_R}$ from $(g_e - 2$) measurements.  The dashed curves show the maximal
  sensitivity of DA$\Phi$NE to $e^+e^- \rightarrow U \gamma$ production, for
  $U \rightarrow e^+ e^-$ (upper, dark green, curve) and $U$ decaying
  invisibly (lower, magenta curve). The DM constraints are essentially
  independent of the ratio $g_\chi / g_{e_R}$, whereas the $g_e-2$ constraint
  as well as the sensitivity limits are independent of $g_\chi$.  Results for
  $g_{e_R} = g_{e_L}$ are similar to those for scalar $\chi$ (Fig.~2).}
\end{figure}

Figs.~1 (2) show the parameter space spanned by $M_U$ and the product of
couplings $g_\chi g_{e_R}$ for the case that $\chi$ is a Majorana fermion
(complex scalar). The regions allowed by the constraints
(\ref{mass}--\ref{omega}) are enclosed by the solid curves, with distinct
allowed regions for $2 m_\chi < M_U$ ($2 m_\chi > M_U$).  We consider only
$g_{e_L}=0$ since this implies $g_{\nu_L} = 0$ if $U$ is a gauge boson of a
gauge group $G_U$ that simply multiplies the gauge group of the Standard Model
(SM). Note that $\nu e$ scattering data imposes the very strong
constraint\cite{fayet}, $g_{\nu_L} \sqrt{g_{e_L}^2 + g_{e_R}^2} < M_U^2 G_F$.
For $g_{e_L} = g_{\nu_L}$ and $g_{e_R}=0$, this would exclude the entire
DM--allowed range (which is invariant under $g_{e_R} \leftrightarrow g_{e_L}$)
of Fig.~1, and most of Fig.~2. Finally, for $g_{e_L}=0$, DM constraints apply
essentially to the product $g_\chi g_{e_R}$. The individual values matter only
if $2 m_\chi \sim M_U$, in which case the value of the decay width $\Gamma_U$
is relevant.

Only a narrow range of couplings is allowed in Fig.~1 on account of an
$S-$wave contribution surviving for our choice of $g_{e_L} = 0$.  Even though
proportional to $m_e^2$, for small $m_\chi$ it would lead to a present
annihilation cross section above the range (\ref{eeann}), unless $2 m_\chi$ is
quite close to $M_U$. In the latter case, the kinetic energy of the DM
particles in the early universe was sufficient to allow efficient annihilation
through the exchange of on--shell $U-$bosons, whereas those at the galactic
center are so slow that they can only annihilate through the exchange of
off--shell $U-$s. This also gives the required large enhancement of the cross
section at decoupling relative to that at the galactic center, even for a
non--vanishing $S-$wave contribution.  However, if $2 m_\chi$ is too close to
(but still below) $M_U$, the relic density constraint (\ref{density}) will
require very small couplings, too small to satisfy the constraint
(\ref{eeann}). For small $M_U$, the allowed range of $m_\chi$ values is,
therefore, very narrow, e.g. 0.72 MeV $\leq m_\chi \leq$ 0.76 MeV for $M_U =
2$ MeV. For larger $M_U$, and correspondingly larger $m_\chi$, the $S-$ wave
contribution $\propto m_e^2$ becomes less important, and a wider range of
values of $m_\chi$ is allowed, e.g. $m_\chi > 11$ MeV for $M_U = 60$ MeV; this
lower bound on $m_\chi$ increases only slowly for even larger values of $M_U$.
Recall that values of $m_\chi$ close to the upper bound of 20 MeV are
preferred~\cite{ahn} since they allow to describe an excess of MeV photons
from the galactic center.

With the $S-$wave contribution vanishing for a scalar $\chi$, the allowed
parameter space is much wider (Fig.~2).  The DM constraints are now compatible
with the entire range of $m_\chi$ except for values very close to $M_U/2$
where today's $\sigma_{\rm ann}$ comes out too small (for $2 m_\chi$ just
below $M_U$) or too large (for $2 m_\chi$ just above $M_U$)~\cite{tiny_range}
if the couplings are chosen to satisfy the constraint (\ref{density}). For
example, for $M_U = 2$ MeV, the range 0.91 MeV $\leq m_\chi \leq$ 1.04 MeV is
excluded. Note that in both Figures, smaller couplings correspond to larger
(smaller) $m_\chi$ if $2 m_\chi < \ (>) M_U$.

Since it is natural to assume $g_{\nu_L}=0$ for our choice of $g_{e_L}=0$, the
only relevant model--independent laboratory constraints come from processes
involving only electrons, the most sensitive being the anomalous magnetic
moment of the electron \cite{fayet}. Using the analytical results of
\cite{leveille} and comparing with the most recent results for SM prediction
and measurement \cite{kinoshita}, we find \cite{bofa}
%
\[
-6 \cdot 10^{-9} \leq \left( \frac {1 \ {\rm MeV}} {M_U} \right)^2 \cdot
   \left( 3 g_{e_L} g_{e_R} - g_{e_L}^2 - g_{e_R}^2 \right) \leq 3 \cdot
   10^{-8}
\]
%
at 95\% C.L.  The resulting upper limit on the product $g_{e_R} \, g_\chi$ is
shown as dotted (blue) line in Figs.~1 and 2. While the data weakly favor a
small positive contribution\cite{boehm2}, $U-$boson loops can account for it
only if $g_{e_L} \simeq g_{e_R}$.  This would either imply a sizable $U \nu
\bar \nu$ coupling, or a model where the $U$ gauge group is embedded
non--trivially into the electroweak gauge group, rendering the construction of
a complete, renormalizable model more complicated.  As long as $g_{e_L}$ is
zero, ($g_e-2$) imposes a rather severe constraint (Figs.1\&2) entirely
independent of $g_\chi$.  With DM-constraints operating essentially on the
product $g_\chi g_{e_R}$, the parameter space that also satisfies the
$(g_e-2)$ constraint becomes larger for larger $g_\chi$.  Together, the
constraints exclude scenarios with $g_\chi \ll g_{e_R}$.  In the opposite
limit, namely $g_\chi = 1 \ (\gg g_{e_R})$, they together exclude scenarios
with $M_U$ much above 200 MeV, as shown in Fig.~2. Other constraints on the
$U-$boson have been discussed in the literature \cite{boehm2,fayet2}; however,
they are more model--dependent in that they involve couplings that are not
required from DM phenomenology. Hence we ignore these constraints here, and
instead turn to a discussion how this model can be tested at $e^+e^-$
colliders. Scenarios with $g_{\nu} = 0$ but $g_e \neq 0$ may also be favored
by BBN~\cite{Serpico:2004nm}.

\begin{figure}[t!]
\vspace*{-5mm}
\rotatebox{270}{\includegraphics[height=0.4\textwidth]{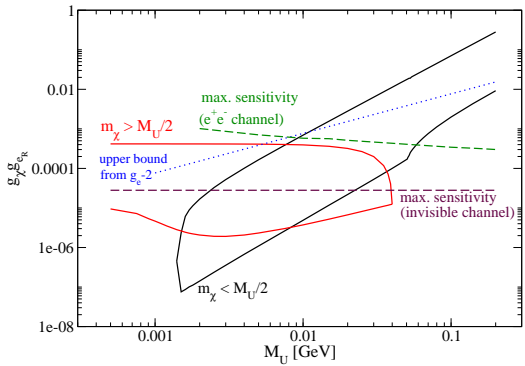}}
\vspace*{-3mm}
\caption{Parameter space of the model with a complex scalar as MeV Dark
  Matter $\chi$ annihilating through the exchange of spin--1 $U$ bosons, for
  $g_{e_L} = g_\nu= 0$ and $g_\chi = 1$. Notation is as in Fig.~1, except that
  the indicated sensitivities are now those that can be achieved at the
  $B-$factories. }
\end{figure}

As noted above, the $U$ boson must couple to $e^+e^-$ pairs in order to
explain the excess flux of 511 keV photons. The process $e^+e^- \rightarrow U
\gamma$ \cite{bofa} will therefore have a non--vanishing cross section if the
cms energy $\sqrt{s} > M_U$:
\begin{equation} \label{signal}
\frac {d \sigma } {d \cos \theta} = \frac
{\alpha \left( g_{e_L}^2 + g_{e_R}^2 \right) } 
{2 s^2 \left( s - M_U^2 \right) } 
\left[ \frac{s^2 + M_U^4}{\sin^2 \theta} -  
            \frac{\left( s - M_U^2 \right)^2}{2}
\right],
\end{equation}
where $\alpha$ is the fine structure constant and $\theta \equiv
\theta_\gamma$ is the emerging angle of the photon. Since existing constraints
imply $M_U \leq 0.2$ GeV, the produced $U$ bosons can decay only into $e^+e^-$
pairs, pairs of DM particles $\chi$, or perhaps neutrinos. This gives rise to
two possible collider signatures, with $e^+e^-\gamma$ and $\gamma +$
``nothing'' final states and we explore the detectability of both.  We focus
on low--energy colliders, since the signal cross section will drop $\propto
1/s$ for $s \gg M_U^2$. Specifically, we analyze the two signatures at the
$\Phi$ factory DA$\Phi$NE, which operates at $\sqrt{s} = m_\phi = 1.02$ GeV,
and at the $B$ factories, which operate at $\sqrt{s} = 10.6$ GeV; the reduced
cross section at the latter is over--compensated by the higher accumulated
luminosity ($\sim 500$ fb$^{-1}$ for both $B-$factories combined, as compared
to $\sim 500$ pb$^{-1}$ at the $\Phi$ factory).

The $e^+e^-\gamma$ final state receives a large contribution from ${\cal
  O}(\alpha^3)$ QED processes. But whereas the signal events have invariant
mass of the outgoing $e^+e^-$ pair $M_{ee}$ very close to $M_U$, the
background distribution has peaks at $M_{ee} \simeq \sqrt{s}$ (from
$t-$channel diagrams with soft $\gamma$ emission) and at a few $m_e$ (from
$s-$channel diagrams).  We therefore require that (i) the produced particles
must not be too close to the beam pipe, $|\cos\theta_i| < 0.9$ for $i = e^\pm,
\, \gamma$; (ii) $M_{ee} \in [M_U - 1 \ {\rm MeV}, \, M_U + 1 \ {\rm MeV}]$,
where the spread is given by the mass resolution of the KLOE detector
\cite{kloe}; we assumed that the BaBar and BELLE detectors at the $B$
factories have similar resolution. The second cut implies that the photon is
quite energetic, since $E_\gamma = (s - M^2_{ee}) / (2 \sqrt{s})$. The signal
is considered detectable if $N_{\rm signal} > 5 \sqrt{N_{\rm bckgd}}$.

The resulting sensitivity limits are indicated by the (upper) dashed (dark
green) curves in Figs.~1 and 2. Signal and background have been calculated
with the CompHEP package \cite{comphep}, augmented to include $U$ bosons.
Note that these are {\em maximal} sensitivities in that we assume a branching
ratio $B(U \rightarrow e^+ e^-) = 1$. This may well be realistic for $2 m_\chi
> M_U$, but for the assumptions made in the Figures is not realistic
otherwise; since we assumed $g_\chi \gg g_{e_R}, \, g_{e_L}$ in these plots,
the invisible $U \rightarrow \chi \bar \chi$ decay mode will dominate if it is
open. However, since the $e^+ e^- \gamma$ final state is background dominated,
the sensitivity limit on $g_{e_R}$ only scales like $B(U \rightarrow
e^+e^-)^{-0.25}$.

Thus, DA$\Phi$NE can probe couplings $g_{e_R}$ down to about $10^{-3}$,
whereas the $B$ factories would be sensitive to couplings as small as $3 \cdot
10^{-4}$. In both cases the sensitivity gets somewhat worse at small $M_U$ as
the background peaks at small $M_{ee}$. Note that these sensitivity limits are
completely independent of $g_\chi$, and of the nature of the DM particle
(scalar or Majorana fermion). In particular, for the scenario of Fig.~1, the
$B$ factories should be able to probe the entire parameter space with $2
m_\chi > M_U$ in this channel. The scenario in Fig.~2 is the worst case
scenario for collider experiments. With the DM constraints essentially fixing
the product of couplings $g_{e_R} g_\chi$ (for $g_{e_L} = 0$), a large
$g_\chi$, as in Fig.~2, therefore leads to small $g_{e_R}$, and hence small
production cross sections.

As already noted, for $g_\chi \gg g_{e_R}$ we expect a large invisible
branching ratio for the $U$ boson if $2 m_\chi < M_U$.  The signal, then,
consists of a single monochromatic photon with $E_\gamma = (s - M_U^2) / (2
\sqrt{s})$. Unfortunately, the experimental resolution on $E_\gamma$ is
considerably worse than that for $M_{ee}$ \cite{kloe}. On the other hand, the
physics (SM)
background now comes from $\nu \bar \nu \gamma$ final states, and is thus
${\cal O}(\alpha G_F^2 s)$. After simple acceptance cuts, $E_\gamma > 100$
MeV, $|\cos \theta| < 0.9$, the background is already completely negligible at
the $\Phi$ factory. Even at the $B$ factories we expect $\ll 1$ background
event once we require $|\cos\theta| < 0.9, \ E_\gamma > 0.5 \sqrt{s} - 200$
MeV~\cite{neut_counting}. In other words, the photon plus ``nothing'' signal
is rate, rather than background, limited. We neglect
instrumental backgrounds here as these are very specific to the experiment.

The corresponding sensitivity limits are shown by the (lower) dashed (maroon)
lines in Figs.~1 and 2. We again show the maximal sensitivity, i.e.  here we
assumed 100\% branching ratio for invisible $U$ decays. We see that
DA$\Phi$NE can probe a coupling $g_{e_R} \gappeq 8 \cdot 10^{-5}$ in this
channel, while the $B$ factories would be sensitive to $g_{e_R} \gappeq 3
\cdot 10^{-5}$, independent of $g_\chi$. In particular, the $B$ factories
would probe the entire parameter space of Fig.~1 with $M_U \geq 4$ MeV in this
channel. Even in the worst case scenario of Fig.~2, much of the DM--allowed
parameter space would lead to a detectable signal.

A dominant invisible decay mode for the $U$ is not realistic if $2 m_\chi >
M_U$, unless $U$ also has significant couplings to neutrinos, which however is
disfavored.  We therefore also investigated $\chi \bar \chi \gamma$
production through off--shell $U$ exchange. The signal cross section is now
proportional to the product $g_{e_R} g_\chi$ (for $g_{e_L}=0$), just like the
DM annihilation cross section (\ref{cctoff}). Using a modified CompHEP, we
find a detectable signal at the $B$-factories ($\geq 10$ events with $|\cos
\theta_\gamma| < 0,9, \ E_\gamma > 0.5 \sqrt{s} - 200$ MeV in 500 fb$^{-1}$)
if $g_{e_R} g_\chi \geq 2.4 \cdot 10^{-4}$. This limit depends only weakly on
$M_U$ and $m_\chi$, so long as $m_\chi$ is not very close to $M_U/2$. This
would be sufficient to probe at least the upper end of the DM allowed region
with $2 m_\chi > M_U$ in Fig.~2.

In summary, we carefully delineated the allowed parameter space of models with
MeV Dark Matter whose annihilation is mediated by the exchange of a spin--1
$U$ boson. Model parameters must be chosen such that the correct thermal relic
density and the correct present annihilation cross section are reproduced; the
latter is motivated by the signal of 511 keV photons from the center of our
galaxy. The parameter space is further constrained by the anomalous magnetic
moment of the electron. We found that these models can be tested decisively at
existing low--energy $e^+e^-$ colliders if the $U$ boson has similar (or
greater) coupling strength to electrons as to DM particles (Fig. 1). 
Such models would
be relatively easy to construct by introducing an additional gauge group with
small coupling constant. Models where the coupling of the $U$ boson to DM
particle is much stronger than that to electrons are much more difficult to
probe at colliders (Fig. 2); 
such a pattern could emerge if $U$ couples to electrons
only through mixing with SM gauge bosons, but has direct coupling to DM
particles. The single photon plus ``nothing'' channel allows to probe much of
the parameter space compatible with the DM constraints even in this
unfavorable situation. We therefore come to the rather surprising conclusion
that the solution of the Dark Matter puzzle might be found at the existing
$\Phi$ and $B$ factories.

{\em Acknowledgments} DC thanks the DST, India for financial assistance and
the Physikalische Institut for hospitality during the inception of this work.
MD thanks the Center for Theoretical Physics at Seoul National University for
hospitality while this work was being completed.


\begin{thebibliography}{9}
\bibitem{tiny_range}
There is a tiny range of $m_\chi$ values very, very close to
  $M_U/2$ where the present cross section is again correct, but it only
  extends over a few eV.
\bibitem{neut_counting}
For a less restrictive cut on $E_\gamma$ we do expect a significant number of
background events at the $B$ factories. In fact, BELLE and BaBar should be
able to determine the number of neutrinos to a few \% using this final
state.
\end{thebibliography}
\end{document}